\documentclass[eqsecnum,aps,superscriptaddress,11pt ]{revtex4}
\usepackage{graphicx}
\usepackage{dcolumn}
\usepackage{xspace}

\begin{document}
\title{Relativistic calculations of the lifetimes and hyperfine structure constants in $^{67}$Zn$^{+}$ }
\vspace*{0.5cm}

\author{Gopal Dixit$^1$, H. S. Nataraj$^2$, B. K. Sahoo$^3$, R. K. Chaudhuri$^2$, and Sonjoy Majumder$^1$ \\
\vspace{0.3cm}
{\it $^1$Department of Physics, Indian Institute of Technology-Madras, Chennai-600 036, India} \\
{\it $^2$ Indian Institute of Astrophysics, Bangalore-34, India}\\
{\it $^3$ Max Planck Institute for the Physics of Complex Systems, D-01187 Dresden, Germany}\\
}
\date{\today}

\begin{abstract}
\noindent
This work presents accurate {\it ab initio} determination of the magnetic dipole (M1) and electric quadrupole 
(E2) hyperfine structure constants for the ground and a few low-lying excited states 
in $^{67}$Zn$^{+}$, which is one of the interesting systems in fundamental physics.
The coupled-cluster (CC) theory within the relativistic framework has been used here in this calculations. Long standing demands for a relativistic and highly
correlated calculations like CC can be able to resolve the disagreements among the lifetime  estimations reported previously for a  
few low-lying states of Zn$^{+}$. The role of different electron correlation effects in the determination of these quantities 
are discussed and their contributions are presented.
\end{abstract}
\maketitle

\section{Introduction}
The quantum information processing (QIP) is one of the interesting areas in physics which is gaining momentum 
both in theoretical and experimental fronts in the recent years. Mostly, the single valence ions particularly 
the ones with $^2S_{1/2}$ ground states are being chosen for QIP studies \citep{ozeri} to encode qubits into the 
hyperfine levels. These levels are chosen due to their relatively long lifetimes against spontaneous 
decay rates and long phase coherence because of their small energy separations. 
The hyperfine structure studies help us understand the nuclear structure of an atom and its influence on the short range 
wavefunctions correctly \cite{bijaya}.

The rapid progress in the 
development of technology involving laser cooling and ion trapping  has made possible to bring these theoretical 
ideas to fruition and singly ionized zinc (Zn$^+$) is one of the recent important inclusion in that family \cite{tanaka}. Though there are a 
few studies of radiative lifetimes of Zn$^+$ in the literature \citep{bergeson}, its hyperfine structures 
are not studied so far limited to the best of our knowledge. Here, we have carried out the magnetic dipole ($A$) and electric quadrupole ($B$) hyperfine 
structure studies of $^{67}$Zn$^+$ for principle quantum number n=4 states which is wanted for the QIP studies as mentioned above. 

Zn is also one of the important elements in astrophysics, especially for the  understanding of the post-main sequence 
evolution of the chemically peculiar stars, in which Zn is either scarce
(if not non-existing) or over abundant \citep{smith}. The high resolution spectra obtained from GHRS onboard 
{\it Hubble Space Telescope} has provided vital informations about its abundances \citep{Savage}. Applications of the radiative
transitions of this ion in cosmology, stellar dynamics, interstellar medium, nucleosynthesis etc. have been discussed 
extensively in the literature \citep{bergeson, smith, astro, popovic}. 

It seems from the reported results that, there exits disagreements in the lifetime estimations among the experimental measurements and various theoretical calculations. 
One distinct feature of the lifetime table is the order of the $4D$ fine structure states, i.e., the lifetime of the $4D_{5/2}$ state
should be less than the lifetime of $4D_{3/2}$ state according to  both experiments available \cite{blagoev, andersen}. This was not found in  
any of the {\it ab initio} studied so far \cite{blagoev, lindgard, curtis, laughlin}. Our calculated lifetimes for the $4D_{3/2}$ and $4D_{5/2}$ states which are reported here, have the same order as well as in 
good agreement with the experimental results.


\section{Theory}

The one-electron reduced matrix elements corresponding to E1, M1 and E2 transitions  are given in
these papers \cite{sahoo04,sahoo06}.
The emission transition probabilities (in sec$^{^-1}$) for the E1, E2 and M1 channels from states {\it f} to {\it i} are given by
\begin{equation}
A^{E1}_{i,f} = \frac{2.0261\times10^{18}}{\lambda^{3}(2j_f+1)}S^{E1},
\end{equation}
\begin{equation}
A^{E2}_{i,f} = \frac{1.11995\times10^{18}}{\lambda^{5}(2j_f+1)}S^{E2},
\end{equation}
\begin{equation}
A^{M1}_{i,f} = \frac{2.69735\times10^{13}}{\lambda^{3}(2j_f+1)}S^{M1},\\
\end{equation}
where $S = {|{\langle \Psi_f|O|\Psi_i\rangle}|}^2$ is the transition strength for the operator O (in a.u.) and $\lambda$ (in(\AA)) is the corresponding transition wavelength. 
The lifetime of a particular state is the reciprocal of the total transition probability arising from
all possible spontaneous electromagnetic transitions from the state to all the lower energy levels. \\
\begin{equation}
\tau_{i} = \frac{1}{A_{i}}. \\
\end{equation}

The interaction between the electromagnetic multipole moments of the electrons and the electromagnetic field created 
at the site of the nucleus is termed as  hyperfine interaction
and the corresponding   Hamiltonian is given by \cite{Cheng}

\begin{equation}
H_{hfs} = \sum_k \bf{M}^{(k)}\cdot \bf{T}^{(k)},
\end{equation}
where $\bf{M}^{(k)}$ and $\bf{T}^{(k)}$ are the spherical tensor operators of
rank $k$ in the nuclear and electronic spaces, respectively. The $k$=1 and 2 terms of the expansion represent the magnetic 
dipole and electric quadrupole interactions, respectively.

The diagonal hyperfine interaction constants can be written as  \cite{Cheng}
\begin{equation}
A = \frac{\mu_I}{IJ} {\langle\gamma JJ|{\bf T^{(1)}_{0}}|\gamma JJ\rangle} = \frac{\mu_I}{IJ} 
\left( \begin{array}{ccc} 
J & 1 & J \\
$-$J & 0  & J \\
\end{array} \right) {\langle J ||{\bf T^{(1)}}||J\rangle},
\end{equation}
and
\begin{equation}
B=2Q{\langle\gamma JJ|{\bf T^{(2)}_{0}}|\gamma JJ\rangle}=2Q 
\left( \begin{array}{ccc}
J & 2 & J \\
$-J$ & 0  & J \\
\end{array} \right){\langle J ||{\bf T^{(2)}}||J\rangle},
\end{equation}
where $I$, $J$ are the total angular momentums of nucleus and electrons; $\mu_I$  and $Q$ are magnetic dipole and 
electric quadrupole moments  of the nucleus, respectively.
The ${\bf T^{(1)}}$ and ${\bf T^{(2)}}$ operators are defined as
\begin{equation}
{\bf T^{(1)}} = \sum_i -ie\sqrt{2}r_i^{-2} {\bf \alpha_i}\cdot C_0^{1}\bigl(\hat{r_i}\bigr)
\end{equation}
and
\begin{equation}
{\bf T^{(2)}} = \sum_i -er_i^{-3} C_q^{2}\bigl(\hat{r_i}\bigr),
\end{equation}
where, $C_q^{k}= \sqrt{4\pi/(2k+1)} Y_{kq}$ with $Y_{kq}$ being the spherical harmonic functions. \\

In the first-order perturbation theory, the hyperfine energy $E_{hfs}(J)$ of the
fine-structure state $|JM_J\rangle$ is the expectation value of the corresponding hyperfine interaction Hamiltonians in that state.
The energies corresponding to the 
magnetic dipole and electric quadrupole hyperfine transition are defined as
\begin{equation}
E_{M1}=AK/2,
\end{equation}
and
\begin{equation}
E_{Q2}={B\over 2}\frac {3K(K+1)-4I(I+1)J(J+1)}{2I(2I-1)2J(2J-1)},
\end{equation}

where $K=2\langle I \cdot J \rangle = F(F+1)-I(I+1)-J(J+1)$ 
with $F = I+J$. Here we have neglected higher order hyperfine interactions.

The basic formalism of the valence universal coupled-cluster (CC) method was developed more than two decades before 
\cite{lindgren,mukherjee,Haque,Pal} however a suitably relativistic version of this approach has been successfully employed 
to obtain the various properties accurately in different single valence atomic systems only recently \cite{kaldor,geetha,majumder,sur,bks}. 
Here we just outline the method applied in this calculation of the wavefunctions of Zn$^+$ accurately. 

The single valence CC theory  extended for the relativistic framework  and is based on the no-virtual-pair approximation 
with Dirac-Fock orbitals \cite{mukherjee}. The concept of the common vacuum for both the closed-shell $N$ and open-shell $N+1$ electron systems
allows to formulate a direct method of excitation energies. 
The dynamical electron correlation effects are introduced through the
{\em valence-universal} wave-operator $\Omega_v$ \cite{mukherjee,lindgren} for the state with $v$ as the valence orbital is written in the normal ordered form
as,
\begin{equation}\label{eq5}
\Omega_v=e^T\{e^{S_v}\}, 
\end{equation}
where cluster operator $T$ represents excitations from the occupied core orbitals of the closed shell system Zn$^{++}$ and $S$ represent 
the core-valence and valence-valence excitations. 
Dominant among these correlations are pair correlations and core polarizations.
The Dirac-Coulomb Hamiltonian dressed with the excitation cluster operators $T$ and $S_v$ are
then diagonalized within the model space constructed from the core and valence orbitals  to obtain the desired
eigenvalues and eigenvectors \cite{Pal}. 
In this work, a leading order
triple excitations are included in the open shell CC amplitude evaluation 
by an approximation
that is similar in spirit to CCSD(T) method \cite{ccsd(t)}. 

The expectation value of any operator $O$ can be expressed in the CC method as
\begin{eqnarray}
O& = & \frac{\langle \Psi_v|O|\Psi_v\rangle}{\langle \Psi_v|\Psi_v\rangle} \nonumber \\
&=& \frac{\langle \Phi_v|\{1+{S_v}^{\dag}\}{e^T}^{\dag}Oe^T\{1+S_v\}|\Phi_v\rangle}
{\langle \Phi_v|\{1+{S_v}^{\dag}\}{e^T}^{\dag}e^T\{1+S_v\}|\Phi_v\rangle}.
\end{eqnarray}

\section{Results and discussions}
We have used  Gaussian-type orbitals (GTO) to calculate the  DF wavefunctions $|\Phi{_{DF}}\rangle$  as given in \cite{rajat} using the
basis functions of the form \cite{mohanty, aerts, visser}\\
\begin{equation}
G_{i,k}(r) = r^{k_{i}}e^{-\alpha_{i}r^2}
\end{equation}
where $k=0, 1, 2, 3,....$ for s, p, d, f ..... type orbital symmetries respectively. 
The large and small components of the relativistic GTOs satisfy the kinetic balance condition \cite{stanton}.
The exponents are determined by the even tempering condition; i.e., for each symmetry exponents are assigned as
\begin{equation}
\alpha_{i} = \alpha_0\beta^{i-1} \hspace{1in} i=1,2,.....N
\end{equation}
where N is the number of basis functions for the specific symmetry.
In this calculation, we have used $\alpha_0=0.00831$ and $\beta=2.99$. 
The number of basis functions used in the present calculation are
32, 32, 30, 25, 20 for $l=$ 0, 1, 2, 3, 4 symmetries, respectively.


\begin{table}[h]
\caption{Radiative lifetimes(ns) for different low-lying states of $^{67}$Zn$^{+}$.}
\begin{ruledtabular}
\begin{tabular}{lrrr}
State &  Experiment & Other theories & This work   \\
\hline
4${^2}P_{1/2}$  &  2.6(3) \cite{blagoev}, 3.05(4) \cite{baumann}, 2.1(4) \cite{hutberg}, 
               & (2.19, 2.01) \cite{blagoev}, 2.151 \cite{lindgard}, (1.97, 2.38) \cite{laughlin},  & 2.43 \\
               &2.5(2) \cite{bergeson}, & 2.7 \cite{migdalck}, 2.524 \cite{curtis}, 2.2 \cite{hutberg}, 3.2 \cite{cowan}, 2.41 \cite{harrison} &  \\
4${^2}P_{3/2}$  &  2.6(3) \cite{blagoev}, 2.07(0.2) \cite{martinson}, 2.1(3) \cite{hutberg}, & (2.1, 1.91) \cite{blagoev}, 2.036 \cite{lindgard}, (1.85, 2.23) \cite{laughlin}, & 2.30\\
              &   3.1(4) \cite{baumann}, 3.0(3) \cite{andersen}, 2.5(2) \cite{bergeson},  & 2.5 \cite{migdalck}, 
2.386 \cite{curtis}, 2.0 \cite{hutberg}, 2.9 \cite{cowan}, 2.27 \cite{harrison}& \\
5${^2}S_{1/2}$  & 1.7(14) \cite{blagoev}, 2.0(2) \cite{Andersen}, 1.8(2) \cite{martinson}, & (1.86, 1.99) \cite{blagoev}, 2.5 \cite{hutberg}, 2.468 \cite{lindgard}, & 2.08 \\
                & 3.85(7) \cite{baumann} & 2.07 \cite{laughlin}, 1.4 \cite{curtis} & \\
4${^2}D_{3/2}$  & 1.8(2) \cite{blagoev}, 1.8(4) \cite{andersen} & (1.27, 1.16) \cite{blagoev}, 2.262 \cite{lindgard}, 1.363 \cite{curtis}, 
1.39 \cite{laughlin} & 1.31\\
4${^2}D_{5/2}$  & 1.44(12) \cite{blagoev}, 1.40(15) \cite{andersen} & (1.31, 1.21) \cite{blagoev}, 1.285 \cite{lindgard}, 1.388 \cite{curtis}, 1.33 \cite{laughlin} & 1.04\\
\end{tabular}
\end{ruledtabular}
\label{tab:results1}
\end{table}

\begin{figure}
\begin{center}
\includegraphics[width=2.5in]{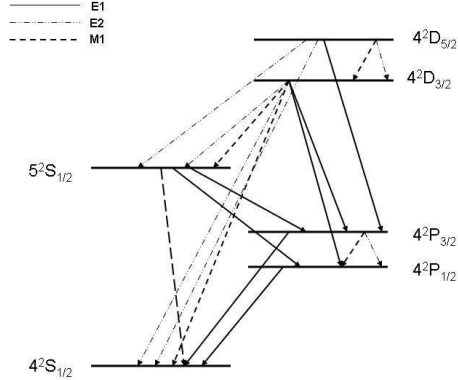}
\caption{\label{label}Decay channels for the first few low-lying excited states of Zn$^+$. The lines of different types correspond to different electromagnetic (multipole) transitions.}
\end{center}
\end{figure}
We report our calculated lifetime results along with the available calculated and measured results in Table I.
It is apparent from the table that, there are large disagreements among the earlier results and also the measurements are not very precise.
In our previous work (here after referred to as paper I \cite{gopal}), we have presented ionization energies, allowed and forbidden transition amplitudes
of the same system considered in the present work and their astrophysical applications are emphasized \cite{gopal}. Our results in paper I are in excellent agreement with the
experimental measurements. With this spirit we have computed the lifetime calculations in the present work which show moderately good agreement with the available experimental results.
There are two recent experiments for the lifetime estimations of the fine structure states
of $4D$ \cite{blagoev, andersen} and both show that, the lifetime of the $4D_{5/2}$ state is shorter than $4D_{3/2}$ state. 
However, many of the earlier theoretical calculations in the literature show the opposite trend, whereas, our CCSD(T) results 
not only show the same trend as that of the experimental results, but also the ratio of their lifetimes  is in an excellent agreement 
with these two measurements. The large transition rate of the $4d_{5/2}$ to $4p_{3/2}$ state (1167893370 $sec^{-1}$) compared to the allowed                
transitions from the $4d_{3/2}$ state to the lower energy states i.e. $4p_{3/2}$ (126252430 $sec^{-1}$) and to $4p_{1/2}$ (642890811 $sec^{-1}$) made this order of lifetime, as seen in Fig. 1.

The computed values of the magnetic dipole hyperfine structure constants ($A$) for the ground state and a few low-lying excited states of $^{67}Zn^{+}$ are given in Table II. Neither the calculations nor the measurements of the hyperfine constants $A$ for all the states, except for the ground state, considered here are available in the literature known to our knowledge. The important many-body correlation contributions to the total $A_{h}$ values are included in our work through relativistic CC theory.
We have used  $\mu_I$ = 0.87547 and $I$ = $5/2$ \cite{raghavan} in our calculation.
From the differences of the DF and the total CC results it is evident that the electron correlation effects
in the calculation of $A$ are quite large and vary from (2-430)$\%$ among the different low-lying states. The core correlation effects are significant inthe $4p_{1/2}$, $4d_{3/2}$ and $4d_{5/2}$ states; especially for the last two states these effects are larger than the DF contributions. The lowest order pair correlation and core polarization effects tabulated here highlight their important contributions which are comparable to the DF contributions, especially to note is the cancellation effect of core polarization in the case of $4p_{3/2}$ state. The large effects of $S_{2v}^{\dagger}\bar{O}S_{2v}$ (4.93 MHz for $4d_{5/2}$ and ~40.0 MHz for $4s$ state)
are observed in these cases.  We have  used the expression (2.10) to calculate the ground state hyperfine energy separation which turns out to be 7018.743 MHz. Panigrahy et al.  also have calculated the same using relativistic linked-cluster many body perturbation theory and get
7.2 GHz \cite{panigrahy} which is in good agreement with our result. The hyperfine energy separation lies in the microwave region of the
electromagnetic spectrum, which suggest that $^{67}$Zn$^{+}$ can be proposed
as the new frequency standard in microwave region, however it needs further investigation about its stability and accuracy of estimation etc.
\\
The computed values of the electric quadrupole hyperfine structure constants ($B$) for a few low-lying excited states are given in Table III. In this calculation the electric quadrupole moment of the nucleus, $Q$ = 0.150 is used \cite{raghavan}. It may be noted that the effects of pair correlation and core polarization effects are stronger than the core correlation effects; in particular, the core polarization effects for the $D$ states are stronger than the DF contributions.\\

\begin{table}
\caption{Magnetic dipole hyperfine constant ($A$) of different low-lying states of
 $^{67}$Zn$^{+}$ in MHz.}
\begin{ruledtabular}
\begin{tabular}{lrrrrrr}
State & DF & Core Correlation & Pair Correlation & Core Polarization & Norm & Total   \\  \hline
4$S_{1/2}$ & 1835.57 & 32.64 & 345.11 & 100.68 & -53.27 & 2339.58   \\
4$P_{1/2}$ & 271.77 & 146.90 & 65.33 & 37.70 & -9.65 & 526.68 \\
4$P_{3/2}$ & 49.94 & 2.07 & 11.68 & -15.84 & -0.92 & 50.74 \\
4$D_{3/2}$ & 6.33 & 6.46 & 0.22 & 1.71 & -0.07 & 15.57\\
4$D_{5/2}$ & 2.72 & 4.54 & 0.72 & 1.57 & -0.07 & 14.56\\
\end{tabular}
\end{ruledtabular}
\label{tab:results1}
\end{table}

\begin{table}
\caption{Electric quadrupole  hyperfine constant ($B$) of different low-lying states of
 $^{67}$Zn$^{+}$ in MHz.}
\begin{ruledtabular}\begin{tabular}{lrrrrrr}
State & DF & Core Correlation & Pair Correlation & Core Polarization & Norm & Total   \\  \hline
4$P_{3/2}$ & 40.01 & 0.72 & 9.43 & 10.74 & -1.12 & 61.74 \\
4$D_{3/2}$ & 1.68 & 0.23 & 0.37 & 2.71 & -0.02 & 5.04\\
4$D_{5/2}$ & 2.40 & -0.22 & 0.48 & 3.15 & -0.03 & 7.35 \\
\end{tabular}    \end{ruledtabular}
\label{tab:results1}
\end{table}


\section{Conclusion}
In this work, we have determined the hyperfine structure constants $A$ and $B$  of the ground state and a few low-lying excited states
in $^{67}$Zn$^+$ using the relativistic coupled-cluster theory.
We have also calculated the hyperfine energy separation for the ground state, which is 7018.743 MHz. There is no experimental result available for the hyperfine energy separation for the ground state, which seems  to be an  important candidate for QIP studies. Also, $^{67}$Zn$^{+}$  can be considered as one of the
promising candidates for the frequency standard in the microwave region. We have also determined the lifetimes of
the low-lying state sin Zn$^{+}$, which are in good agreement with experimental results. Especially our calculated lifetimes of 4$D$ fine structure states explain the same trend as observed in the experiments \cite{blagoev, andersen} i.e., the lifetime of $4D_{5/2}$ state is shorter than the $4D_{3/2}$ state, unlike many other
theoretical results which show opposite trend and also the ratio of their lifetimes in our calculation is in excellent agreement with the experimental result.This suggests that the relativistic CC method applied in the present work and our numerical approach in obtaining the wavefunctions of the system considered are more accurate and reliable.

\section{Acknowledgment}
We greatly acknowledge Prof. B. P. Das, Indian institute of Astrophysics, Bangalore and Prof. Debashis Mukherjee, Indian Association of Cultivation for Science, Kolkata for the helpful discussions.

\end{document}